\documentclass[11pt]{kimreview}

\usepackage{bm}
\usepackage{amsmath}
\usepackage{amssymb}
\usepackage{epstopdf} 
\usepackage{wrapfig}  
\usepackage{lipsum}   
\usepackage{cite}

\DeclareUnicodeCharacter{2212}{-}
\begin{document}


\title{Boltzmann Generators and the New Frontier of Computational Sampling in Many-Body Systems}

\authorone{Alessandro Coretti}
\affiliationone{Faculty of Physics, University of Vienna, Austria}
\authortwo{Sebastian Falkner}
\affiliationtwo{Faculty of Physics, University of Vienna, Austria}
\authorthree{Jan Weinreich}
\affiliationthree{Faculty of Physics, University of Vienna, Austria}
\authorfour{Christoph Dellago}
\affiliationfour{Faculty of Physics, University of Vienna, Austria}
\authorfive{O. Anatole von Lilienfeld}
\affiliationfive{Vector Institute; Departments of Chemistry and Materials Science and Engineering, University of Toronto, Canada}



\publishyear{2024}
\volumenumber{2}
\articlenumber{03}
\submitdate{November 29, 2023}
\publishdate{April 22, 2024}
\doiindex{10.25950/bfa99422}
\doilink{10.25950/bfa99422}


\paperReviewed
{Boltzmann generators: Sampling equilibrium states of many-body systems with deep learning}
{Frank Noé, \textit{et al}.}
{\href{https://doi.org/10.1126/science.aaw1147}{Science, 365:6457 (2019)}}


\maketitle


\begin{abstract}
The paper by Noé \textit{et al}.\cite{bib_1} introduced the concept of Boltzmann Generators (BGs), a deep generative model that can produce unbiased independent samples of many-body systems. They can generate equilibrium configurations from different metastable states, compute relative stabilities between different structures of proteins or other organic molecules, and discover new states.
In this commentary, we motivate the necessity for a new generation of sampling methods beyond molecular dynamics, explain the methodology, and give our perspective on the future role of BGs.
\end{abstract}
\medskip

\section*{Background}

Before delving into the discussion of the paper by Noé \textit{et al}.\cite{bib_1}, it is essential to first outline the main challenge it seeks to address. Within the domain of numerical atomistic simulations, two significant issues frequently dominate computational complexity: the first is the computational “curse” of solving the electronic Schrödinger equation, prohibiting chemically accurate \textit{first principles} investigations of large molecules. The second is the so-called sampling problem: Even when using predictive machine learned potentials, i.e. data-driven and cost-effective approximations of the electronic potential, or more conventional force fields, it is impossible to reach the timescales necessary for many chemical and biological processes. While machine learned energies\cite{bib_2}, or forces\cite{bib_3, bib_4, bib_5, bib_6} recover even highly accurate quantum labels orders of magnitude faster than numerical solutions of approximate variants of Schrödinger’s equation, they can still be substantially slower than traditional force fields\cite{bib_7, bib_8, bib_9, bib_10}. Furthermore, the sampling problem of \textit{uncorrelated} physical configurations within statistical mechanics ensembles remains. This latter problem is closely intertwined with computing free energies that govern the phase diagrams of condensed matter. To achieve that, sufficient coverage of uncorrelated configurations must be guaranteed. However, the challenge is that directly simulating trajectories of the many atoms that make up materials and molecules in order to integrate Newton’s equations of motion, and to compute their essential properties for most relevant time scales, is computationally prohibitive, exceeding even the capabilities of supercomputers.

Using generative deep neural networks, Noé \textit{et al}. tackled the sampling problem from a novel direction in 2019. From their first appearance, the use of generative neural networks has been tempting in statistical mechanics, due to their ability to produce independent samples from a given distribution. Provided sufficient training data, this could effectively overcome some of the most challenging problems of standard sampling algorithms, commonly used in statistical mechanics, in particular correlations between subsequently sampled states. Generative models have originally been developed in the realm of image/text/audio generation, where a large set of examples is available and no analytical form of the target distribution exists. Within the realm of the atomistic sciences, they were introduced for the purpose of molecular materials design already one year earlier in 2018\cite{bib_11}. The physical sampling problem, however, is profoundly different since the exact target distribution is known (up to a proportionality constant) and since it is crucial to sample the target distribution exactly, to avoid any bias in the result of the numerical simulation. The contribution of Noé \textit{et al}. was also contextualized in the same issue within a perspective by Tuckerman\cite{bib_12}.

Normalizing flows are a particular class of deep generative models well suited to accommodate these different premises. First, they can be trained by exploiting the analytical likelihood of the target space to sample from. Second, the architecture of the network allows to analytically compute the likelihood of a generated sample. This allows for the generation of a fully unbiased distribution in target space. These two features make normalizing flows a very promising tool for tackling the sampling problem of physical configurations. Incidentally, related work was published two weeks earlier in \textit{Phys Rev D} on flow-based generative models for Markov chain Monte Carlo in lattice field theory\cite{bib_13}. Corresponding successful applications have been reported more recently,\cite{bib_14, bib_15, bib_16, bib_17} and also for the calculation of free energies\cite{bib_18, bib_19}. In their paper, Noé \textit{et al}. have adapted the architecture of normalizing flows to the physical sampling problem, using the energy of the target system as the likelihood for training the model. They then introduce Boltzmann generators (BGs), which are normalizing flows aimed at solving the sampling problem of statistical mechanics.

\section*{Summary of the article}

The architecture of BGs is built on normalizing flows: The idea is to train a deep neural network to approximate a transformation from a “latent” space $\mathrm{z}$ to a target space $\mathrm{x}$ such that

\begin{equation}
    x=F_{\mathrm{zx}}(z)
    \label{eqn_1}
\end{equation}
where $z$ and $x$ are samples from the spaces $\mathrm{z}$ and $\mathrm{x}$ with distributions $\mu_\mathrm{z}(z)$ and $\mu_\mathrm{x}(x)$, respectively. The distribution of the latent space, which is sometimes called “prior”, is usually chosen to be very simple (Gaussian or even uniform). In this way, once the network is trained, it is possible to easily get samples from $\mathrm{z}$ and to transform them, via $F_{\mathrm{zx}}$, to samples distributed according to $\mu_\mathrm{x}$. Choosing the target distribution $\mu_\mathrm{x}(x) = e^{ − \beta U(x)}$, the BG samples the NVT ensemble of the system defined by the potential energy $U(x)$ at inverse temperature, $\beta = (k_BT)^{−1}$.

The neural network representing the transformation $F$ is built to be invertible. In this way, one can also compute $z = F_{\mathrm{xz}}(x)$ where $F_{\mathrm{xz}} = F_{\mathrm{zx}}^{−1}$. The invertibility of the transformation $F$ guarantees that, given the likelihood of a sample in latent space $\mu_\mathrm{z}(z)$, it is possible to exactly compute the corresponding likelihood in the target space after the transformation as $p_{\mathrm{x}}(x) = \mu_\mathrm{z}(z) \det |J_{\mathrm{zx}}(z)|^{−1}$ where $J_{\mathrm{zx}}$ is the Jacobian matrix associated to $F_{\mathrm{zx}}$. The same is true for the inverse transformation which yields $p_{\mathrm{z}}(z) = \mu_\mathrm{x}(x) \det |J_{\mathrm{xz}}(x)|^{−1}$. It is important to highlight the difference between the distributions $\mu$ and $p$. The former corresponds to the \textit{true} distribution of samples in space, while the latter is the distribution of samples that is generated by the network. The two will be different in general.

The goal of making the target distributions $\mu_\mathrm{x}$ and $p_\mathrm{x}$ as close as possible provides a natural way of training the network. The Kullback-Liebler divergence as a “distance” in distribution space provides the loss function
\begin{equation}
J_{\mathrm{KL}}=\mathbb{E}_\mathrm{z}\bigl[\beta U(F_\mathrm{zx}(z))−\log(\det J_\mathrm{zx}(z))\bigr]
\label{eqn_2}
\end{equation}
where $\mathbb{E}_\mathrm{z}$ is the mean value over a batch of samples from $\mathrm{z}$. This loss function corresponds to the free-energy difference between the prior and the target distributions. Minimizing Eq.~(\ref{eqn_2}) corresponds to a) minimizing the internal energy (first term) and therefore training the network to sample low-energy configurations, and b) maximizing the entropy of the target distribution at the given temperature (second term) and therefore avoiding mode-collapse in the lowest energy configuration. The invertibility of the network also allows for training in the other direction. This proves to be particularly useful to initialize the flow at the beginning of the training process. Given a subset of initial configurations from the target space sampled by means of standard algorithms such as Molecular Dynamics (MD) or Monte Carlo (MC), one can maximize the likelihood in the distribution of latent space (from here on assumed to be normal) of samples transformed via $F_\mathrm{xz}(x)$. This yields the loss function
\begin{equation}
    J_{\mathrm{ML}} = \mathbb{E}_\mathrm{x} \biggl[\frac{1}{2} \|F_\mathrm{xz}(x) \|^2 - \log(\det J_\mathrm{xz}(x)) \biggr]
    \label{eqn_3}
\end{equation}
This second loss function is crucial when multiple minima are present in the potential energy surface of the target system. In this case, the entropic term in Eq.~(\ref{eqn_2}) alone is not sufficient to avoid the collapse of the generated configurations around a single minimum. Eqs.~(\ref{eqn_2}) and~(\ref{eqn_3}) represent the two main terms of the total loss function used for training and they are in practice often combined in a single loss function (sometimes with different weights). Also additional, system-specific terms can be added at convenience. In particular, the authors discuss a Reaction Coordinate (RC) loss that can be optionally included to force the system to generate configurations closer to energy barriers.

Finally, knowledge of the likelihood of a transformed sample allows for the removal of any residual bias in the generated distribution. For example, a weight $\omega(x) = \mu_\mathrm{x}(x)/p_\mathrm{x}(x)$ can be assigned to every generated configuration $x$ and statistical mechanics estimators can be computed as
\begin{equation}
    A = \frac{\Sigma_i \omega(x_i)a(x_i)}{\Sigma_i\omega(x_i)}
    \label{eqn_4}
\end{equation}
where $A$ is an unbiased estimator of an observable $\langle A \rangle$.

Given this basic architecture, the authors also provide an algorithm to explore the target space while training the BG. Starting from a pool of known physical configurations, they perform a MC simulation in latent space and progressively add more configurations to the initial pool. Training the flow with the newly added configurations as the exploration proceeds, guarantees that the mapping to the latent space is always consistent with the currently explored physical space. The acceptance criterion of the MC scheme ensures that the generated samples approach the correct Boltzmann distribution.

The authors show applications of the proposed methods to a wide range of cases, from proof-of-principle calculations for systems with two degrees of freedom to complex biochemical systems such as bovine pancreatic trypsin inhibitor (BPTI) protein, passing through a two dimensional example of a condensed matter system.

\section*{Commentary and critical analysis}

From the dawn of molecular simulations in the 1950s, the introduction of BGs represents one of the few attempts at a paradigm shift in the calculation of statistical mechanics observables via numerical experiments. Most of the efforts in the development of methods for generating physical configurations are focused on the improvement of the Markov chain upon which the methods are based. Enhanced sampling algorithms, replica exchange methods, and multi-scale techniques (just to name a few) allowed tremendous advances in the field, but none of these method drastically changed the way in which sampling is carried out, i.e. via sequential updates of configurations. Boltzmann generators tackle the sampling problem from a different angle, making the most out of the increase in computational power provided by the latest advances in deep learning.

Due to this change of perspective, assessing the performances of BGs with respect to standard sampling algorithms is not straightforward. The generation of configurations via BGs is astonishingly fast. The time needed to transform samples from latent space is many orders of magnitude lower than the time needed for producing the same number of independent configurations using standard Markov-chain-based methods. This comes at the cost of a painful and time-consuming training process, which involves fine-tuning different hyperparameters and a large number of energy evaluations. Even if it is true that the training process is a one-time procedure, it is also true that, for the time being, there is only a limited transferability from system to system. In practice, it has been observed that the performance gain obtained using BGs tends to be strongly dependent on the system under investigation. The training process is complicated by complex energy landscapes and a large number of degrees of freedom as this task usually requires bigger networks and larger training sets. On the other hand, strong correlations between sampled configurations and the presence of high energy barriers and basins of attraction in configuration space can become problematic for standard methods and time needed to decorrelate or to “jump” out of such basins can quickly amortize for the training time.

Some limitations in the applicability of the methods also arise from the relatively new architecture of normalizing flows underpinning BGs. The network's invertibility prevents a significant reduction in the number of degrees of freedom that the transformation must act upon. The treatment of large systems is therefore hampered by the inherently global nature of the transformation which limits its expressiveness. This can be particularly limiting for biochemical and biological systems in which solvation effects are important and for which good models of implicit solvents are not available. A high number of degrees of freedom is also expected to have an impact on the difficulty of the training process and on the reweighting algorithm. The latter, in particular, to be effective requires a degree of superposition between the target and generated distributions which scales exponentially with the number of degrees of freedom.

Our experience\cite{bib_20,bib_21} has also shown that increasing the “power” of the network, i.e. the ability to faithfully reproduce a given complex distribution, is not so straightforward as in other neural network applications. Oftentimes, the mantra \textit{bigger is better} does not apply to normalizing flows, and increasing the network size has only a mild effect or indeed no effect on the quality of the training. What seems to be effective in increasing the efficiency of the training is a different and somewhat smarter representation of the input. A very good example is represented by the mixed-coordinate transformation the authors used to treat the BPTI protein. The designing of smart transformation layers (usually non-learnable layers that are placed between the raw input, e.g. the 3$N$-dimensional array containing the particle coordinates, and the normalizing flow), can make a difference in the training process.

While BGs, for the time being, are still too “young” to provide a comparable alternative to standard sampling methods, in particular for big and complex systems, we see the enormous potential that the introduction of BGs represents for the molecular simulation community. On the other hand, the use of BGs in conjunction with other standard sampling algorithms, in the spirit of the exploration method proposed by the authors, can already drastically improve the efficiency of molecular simulations today. One clear example is the use of BGs to propose smart Monte Carlo moves\cite{bib_22,bib_23}.

Further progress in the field of generative models is expected to mitigate many of the remaining issues of the method and eventually make it the new standard for generating equilibrium configurations of statistical mechanics systems.

\section*{Potential new directions}

Many research directions stemmed, directly or indirectly, from the publication of BGs in \cite{bib_1} and from the efforts of the community to improve on the baseline provided by this paper. Some of these efforts are discussed and referenced below.

Soon after the publication of \cite{bib_1}, the same group explored different techniques that could improve on previous results. For example, in Ref. 24 they designed a BG which could be trained to sample exactly many thermodynamics states with a single training. This was achieved by parameterizing the generated distribution through the temperature of the target ensemble. To this end, they employed a different architecture with respect to the original paper: they augmented the physical space by doubling the degrees of freedom involved in the transformation and they designed the architecture of the network to be volume preserving, (i.e. $\det|J|=1$).

In a different paper\cite{bib_25} they introduced stochastic dynamics between the deterministic blocks of the normalizing flow. They show how to overcome, in this way, possible topological constraints in the target density and increase the expressiveness of the network in reproducing complex target distributions. Moreover, they show how the combined optimization of the network parameters and of the stochastic sampler also improves the efficiency of the latter. This approach has been pursued also by other groups in different fields of statistical mechanics\cite{bib_26, bib_27}.

The problem of topological constraints is tackled by a different perspective in a later paper\cite{bib_28}, where the condition of the smoothness of the flow is also preserved and the gradient computation is also addressed for these cases. In the same paper, the possibility of using forces in the target space $\mathcal{F}(x)$ to improve training is also introduced. The authors show how the addition in the loss function of a Force Matching (FM) term

\begin{equation}
J_{\mathrm{FM}}=\mathbb{E}_\mathrm{x}
\biggl[
\|\mathcal{F}(x) - \nabla_x \log p_\mathrm{x} (x)\|^2 
\biggr] \label{eqn_5}
\end{equation}
can drastically improve the training efficiency while compensating for the additional burden of computing force on data in target space.

In a more recent paper\cite{bib_29}, the power of continuous normalizing flows\cite{bib_30,bib_31} has been combined with the simplified training process arising from flow matching\cite{bib_32,bib_33} and with the possibility of encoding some of the symmetries of the system in the transformation itself via the design of equivariant flows\cite{bib_34, bib_35}. Rotational symmetries are also discussed using quaternions in Ref. 36.

The training process can be facilitated by the introduction of physical information in the prior distribution. Wirnsberger \textit{et al}.\cite{bib_37} generated configurations for a solid using normalizing flows, relying on a prior distribution that consisted in Gaussians placed at the lattice sites of the crystalline phase of the target system. In this way, they managed to generate very accurate configurations with relatively little effort and without the need for reweighting. Moreover, they managed to achieve this result with no reference structures from the target system. Coretti \textit{et al}.\cite{bib_21} followed a similar line of research for liquid systems, generating liquid configurations using simulations with a simpler potential energy at a higher temperature as the prior distribution.

By integrating physical priors about the intrinsic distribution of internal coordinates, new approaches could capitalize on a) the inherently Gaussian characteristics of bonded interactions derived from the harmonic potential of bonds, and b) the specific distributions of non-bonded atom pairs. Embedding such physical priors to intelligently parameterize the learning model could reduce it to just a handful of parameters for each type of interaction. For example, in Ref. 38, a model was trained to predict the average geometry of small molecules at finite temperatures using a single machine learning model for each bond pair.

The use of BGs is also promising in relation to the study of rare events. Falkner \textit{et al}.\cite{bib_20} produced a version of BGs that can be conditioned to generate configurations biased along a reaction coordinate in target space. They show how this is very powerful either for computing free energy or for producing shooting points for transition path algorithms.

Statistical ensembles other than NVT have also been investigated by different research groups: Wirnsberger \textit{et al}.\cite{bib_39} and van Leeuwen \textit{et al}.\cite{bib_40}, almost at the same time, came out with an algorithm for the generation of configurations in the isobaric-isothermal ensemble whose main features are based on BGs.

Finally, a word on research directions that have not yet been addressed but where normalizing flows could make a positive impact: In situations where an analytical expression for the nonequilibrium distribution is present, the use of BGs to sample nonequilibrium configurations could be of interest to many applications. For instance, Quantum Monte Carlo or Path Integrals could also be among the applications that could benefit from machines which generate one-shot configurations out of a given distribution\cite{bib_41}.

\section*{Acknowledgment}

O.A.v.L. has received support as the Ed Clark Chair of Advanced Materials and as a Canada CIFAR AI Chair. A.C., S.F. and C.D. acknowledge support of the Austrian Science Fund (FWF) through the SFB TACO, grant number F 81-N.



\begin{thebibliography}{9}

\bibitem{bib_1}
Noé, F., Olsson, S., Köhler, J. and Wu, H., Boltzmann generators: Sampling equilibrium states of many-body systems with deep learning, \emph{Science}, 365:eaaw1147 (2019).
\bibitem{bib_2}
Huang, B. and von Lilienfeld, O. A., Ab initio machine learning in chemical compound space, \emph{Chemical Reviews}, 121:10001 (2021).
\bibitem{bib_3}
Bartók, A., Csányi, G., Gaussian approximation potentials: A brief tutorial introduction, \emph{International Journal of Quantum Chemistry}, 115:1051 (2015).
\bibitem{bib_4}
Behler, J. Four Generations of High-Dimensional Neural Network Potentials, \emph{Chemical Reviews}, 121:10037–10072 (2021).
\bibitem{bib_5}
Unke, O. T., Chmiela, S., Sauceda, H. E., Gastegger, M., Poltavsky, I., Schütt, K. T., Tkatchenko, A., Müller, K.-R. Machine Learning Force Fields, \emph{Chemical Reviews}, 121:10142–10186 (2021).
\bibitem{bib_6}
Käser, S., Vazquez-Salazar, L. I., Meuwly, M., Töpfer, K. Neural network potentials for chemistry: concepts, applications and prospects, \emph{Digital Discovery}, 2:28–58 (2023).
\bibitem{bib_7}
Wang, J., Wang, W., Kollman, P. A., Case, D. A. Automatic atom type and bond type perception in molecular mechanical calculations, \emph{J. Mol. Graph. Model.}, 25:247–260 (2006).
\bibitem{bib_8}
Wang, J., Wolf, R. M., Caldwell, J. W., Kollman, P. A. Development and testing of a general amber force field, \emph{J. Comp. Chem.}, 25:1157–1174 (2004).
\bibitem{bib_9}
Qiu, Y. et al. Development and Benchmarking of Open Force Field v1.0.0—the Parsley Small-Molecule Force Field, \emph{Journal of Chemical Theory and Computation}, 17:6262–6280 (2021).
\bibitem{bib_10}
 Bjelkmar, P., Larsson, P., Cuendet, M. A., Hess, B., Lindahl, E. Implementation of the CHARMM Force Field in GROMACS: Analysis of Protein Stability Effects from Correction Maps, Virtual Interaction Sites, and Water Models, \emph{Journal of Chemical Theory and Computation}, 6:459–466 (2010).
\bibitem{bib_11}
Gómez-Bombarelli, R., Wei, J. N., Duvenaud, D., Hernández-Lobato, J. M., Sánchez-Lengeling, B., Sheberla, D., Aguilera-Iparraguirre, J., Hirzel, T. D., Adams, R. P., Aspuru-Guzik, A. Automatic chemical design using a data-driven continuous representation of molecules, \emph{ACS central science}, 4:268–276 (2018).
\bibitem{bib_12}
Tuckerman, M. E. Machine learning transforms how microstates are sampled, \emph{Science}, 365:982–983 (2019).
\bibitem{bib_13}
Albergo, M. S., Kanwar, G., Shanahan, P. E. Flow-based generative models for Markov chain Monte Carlo in lattice field theory, \emph{Physical Review D}, 100:034515 (2019).
\bibitem{bib_14}
Nicoli, K. A., Anders, C. J., Funcke, L., Hartung, T., Jansen, K., Kessel, P., Nakajima, S., Stornati, P. Estimation of thermodynamic observables in lattice field theories with deep generative models, \emph{Physical review letters}, 126:032001 (2021).
\bibitem{bib_15}
Nicoli, K. A., Nakajima, S., Strodthoff, N., Samek, W., Müller, K.-R., Kessel, P. Asymptotically unbiased estimation of physical observables with neural samplers, \emph{Physical Review E}, 101:023304 (2020).
\bibitem{bib_16}
Singha, A., Chakrabarti, D., Arora, V. Conditional normalizing flow for Markov chain Monte Carlo sampling in the critical region of lattice field theory, \emph{Physical Review D}, 107:014512 (2023).
\bibitem{bib_17}
Gabrié, M., Rotskoff, G. M., Vanden-Eijnden, E. Adaptive Monte Carlo augmented with normalizing flows, \emph{Proceedings of the National Academy of Sciences}, 119:e2109420119 (2022).
\bibitem{bib_18}
Ahmad, R., Cai, W. Free energy calculation of crystalline solids using normalizing flows, \emph{Modelling and Simulation in Materials Science and Engineering}, 30:065007 (2022).
\bibitem{bib_19}
Wirnsberger, P., Ballard, A. J., Papamakarios, G., Abercrombie, S., Racaniére, S., Pritzel, A., Jimenez Rezende, D., Blundell, C. Targeted free energy estimation via learned mappings, \emph{The Journal of Chemical Physics}, 153 (2020).
\bibitem{bib_20}
Falkner, S., Coretti, A., Romano, S., Geissler, P. L., Dellago, C. Conditioning Boltzmann generators for rare event sampling, \emph{Machine Learning: Science and Technology}, 4:035050 (2023).
\bibitem{bib_21}
Coretti, A., Falkner, S., Geissler, P., Dellago, C. Learning mappings between equilibrium states of liquid systems using normalizing flows, arXiv preprint, arXiv:2208.10420 (2022).
\bibitem{bib_22}
Sbailò, L., Dibak, M., Noé, F. Neural mode jump monte carlo, \emph{The Journal of Chemical Physics}, 154 (2021).
\bibitem{bib_23}
Invernizzi, M., Krämer, A., Clementi, C., Noé, F. Skipping the replica exchange ladder with normalizing flows, \emph{The Journal of Physical Chemistry Letters}, 13:11643–11649 (2022).
\bibitem{bib_24}
Dibak, M., Klein, L., Krämer, A., Noé, F. Temperature steerable flows and Boltzmann generators, \emph{Phys. Rev. Res.}, 4:L042005 (2022).
\bibitem{bib_25}
Wu, H., Köhler, J., Noé, F. Stochastic normalizing flows, \emph{Advances in Neural Information Processing Systems}, 33:5933–5944 (2020).
\bibitem{bib_26}
Caselle, M., Cellini, E., Nada, A., Panero, M. Stochastic normalizing flows for lattice field theory, arXiv preprint, arXiv:2210.03139 (2022).
\bibitem{bib_27}
Caselle, M., Cellini, E., Nada, A., Panero, M. Stochastic normalizing flows as non-equilibrium transformations, \emph{Journal of High Energy Physics}, 1–31 (2022).
\bibitem{bib_28}
Köhler, J., Krämer, A., Noé, F. Smooth normalizing flows, \emph{Advances in Neural Information Processing Systems}, 34:2796–2809 (2021).
\bibitem{bib_29}
Klein, L., Krämer, A., Noé, F. Equivariant flow matching, arXiv preprint, arXiv:2306.15030 (2023).
\bibitem{bib_30}
Chen, R. T., Rubanova, Y., Bettencourt, J., Duvenaud, D. K., Neural ordinary differential equations, \emph{Advances in neural information processing systems}, 31 (2018).
\bibitem{bib_31}
Grathwohl, W., Chen, R. T., Bettencourt, J., Duvenaud, D., Scalable reversible generative models with free-form continuous dynamics, \emph{International Conference on Learning Representations} (2019) p 7.
\bibitem{bib_32}
Lipman, Y., Chen, R. T., Ben-Hamu, H., Nickel, M., Le, M., Flow matching for generative modeling, arXiv preprint, arXiv:2210.02747 (2022).
\bibitem{bib_33}
Tong, A., Malkin, N., Huguet, G., Zhang, Y., Rector-Brooks, J., Fatras, K., Wolf, G., Bengio, Y., Conditional flow matching: Simulation-free dynamic optimal transport, arXiv preprint arXiv:2302.00482 (2023).
\bibitem{bib_34}
Rezende, D. J., Racanière, S., Higgins, I., Toth, P. Equivariant hamiltonian flows, arXiv preprint, arXiv:1909.13739 (2019).
\bibitem{bib_35}
Köhler, J., Klein, L., Noé, F., Equivariant flows: exact likelihood generative learning for symmetric densities, \emph{International conference on machine learning} (2020) pp 5361–5370.
\bibitem{bib_36}
Köhler, J., Invernizzi, M., De Haan, P., Noé, F. Rigid body flows for sampling molecular crystal structures, arXiv preprint, arXiv:2301.11355 (2023).
\bibitem{bib_37}
Wirnsberger, P., Papamakarios, G., Ibarz, B., Racanière, S., Ballard, A. J., Pritzel, A., Blundell, C. Normalizing flows for atomic solids, \emph{Machine Learning: Science and Technology}, 3:025009 (2022).
\bibitem{bib_38}
Weinreich, J., Lemm, D., von Rudorff, G. F., von Lilienfeld, O. A. Ab initio machine learning of phase space averages, \emph{The Journal of Chemical Physics}, 157:024303 (2022).
\bibitem{bib_39}
Wirnsberger, P., Ibarz, B., Papamakarios, G. Estimating Gibbs free energies via isobaric-isothermal flows, \emph{Machine Learning: Science and Technology}, 4:035039 (2023).
\bibitem{bib_40}
    van Leeuwen, S. and Ortíz, A. P. d. A. and Dijkstra, M. A, Boltzmann generator for the isobaric-isothermal ensemble, arXiv preprint, arXiv:2305.08483 (2023).
\bibitem{bib_41}
    Tuckerman, M. E. Statistical Mechanics: Theory and Molecular Simulation, \emph{Oxford University Press} (2010), p 712.

\end{thebibliography}
\end{document}